\documentclass[twocolumn,english,aps,prb,groupedaddress,showpacs,floatfix]{revtex4}

\usepackage[utf8]{inputenc}  
\usepackage[active]{srcltx}  

\usepackage{color}
\usepackage{epsfig}
\usepackage{graphicx}
\usepackage{pstricks}

\usepackage{amsmath}
\usepackage{amssymb}
\usepackage{amsfonts}
\usepackage{amsthm}
\usepackage{bm}

\makeatletter
\@ifundefined{textcolor}{}
{%
 \definecolor{BLACK}{gray}{0}
 \definecolor{gray}{gray}{0.5}
 \definecolor{WHITE}{gray}{1}
 \definecolor{RED}{rgb}{1,0,0}
 \definecolor{GREEN}{rgb}{0,1,0}
 \definecolor{BLUE}{rgb}{0,0,1}
 \definecolor{CYAN}{cmyk}{1,0,0,0}
 \definecolor{MAGENTA}{cmyk}{0,1,0,0}
 \definecolor{YELLOW}{cmyk}{0,0,1,0}
}
\makeatother

\begin{document}


\title{Topological solitons and bulk polarization switch in collinear type II multiferroics}

\author{D.\ C.\ Cabra}

\affiliation{IFLySiB-CONICET and Departamento de F\'{i}sica, Universidad Nacional
de La Plata, 
Argentina}

\author{A.\ O.\ Dobry}

\affiliation{IFIR-CONICET and Facultad de Ciencias Exactas, Ingeniería y Agrimensura,
Universidad Nacional de Rosario, Argentina}

\author{C.\ J.\ Gazza }

\affiliation{IFIR-CONICET and Facultad de Ciencias Exactas, Ingeniería y Agrimensura,
Universidad Nacional de Rosario, Argentina}

\author{G.\ L.\ Rossini }

\affiliation{IFLP-CONICET and Departamento de F\'{i}sica, Universidad Nacional
de La Plata, 
Argentina}

\date{\today}

\begin{abstract}
	We introduce a microscopic model for collinear multiferroics
	capable to reproduce, 
	as a consequence of magnetic frustration and easy-axis anisotropy, 
	the so-called ``uudd" (or antiphase) magnetic ordering observed in several type II multiferroic materials.
	The crucial role of lattice distortions in the multiferroic character of these materials 
	is entered into the model via an indirect magnetoelectric coupling, 
	mediated by elastic degrees of freedom through a pantograph mechanism.
	Long range dipolar interactions set electric dipoles in the antiferroelectric order.
	We investigate this model by means of extensive DMRG computations and complementary analytical methods.
	We show that a lattice dimerization induces an spontaneous $\mathbb{Z}_2$ ferrielectric bulk polarization,
	with a sharp switch off produced by a magnetic field above a critical value.
	The topological character of the magnetic excitations makes this mechanism robust.
	%


\end{abstract}


\maketitle

\section{Introduction\label{sec:Introduction}}
Multiferroic materials, defined as those in which ferroelectricity and (anti)ferromagnetism coexist and interact, 
have become one of the most studied topics in the last few years, 
both from the experimental and the theoretical point of view. 
The possibility of magnetic writing via electric fields makes these materials a potential source of 
technological applications in data storage. 

Among the most recent discoveries a 
type of magnetoelectric materials, so called type II multiferroics 
in which the electrical polarization coincides with a magnetic ordering transition, 
has been the subject of a lot of efforts [\onlinecite{CheongMostovoy, Khomskii}].
What is most important in these materials is the very large coupling between magnetic and electrical properties,
even if the value of the electrical polarization can be rather small as compared to typical ferroelectric materials.

An important issue is to determine the microscopic underlying general mechanism which could be applied 
to guide the synthesis of bulk or film materials with enhanced magnetoelectric properties
(see \textit{e.g.} [\onlinecite{Radaelli-2008}] and references therein). 
Within the present paper we contribute to 
this task
by proposing and analyzing a model in which this cross-coupling arises 
via the interaction with the lattice, 
thus fitting into the so called exchange-striction mechanism [\onlinecite{CheongMostovoy,Khomskii,Dagotto-2019}]

Very generally the high magnetoelectric response appears to be associated to the magnetic frustration due to 
competing spin interactions leading to complex magnetic orders [\onlinecite{CheongMostovoy}].
Indeed, in most of multiferroic materials with  collinear spins
the magnetic order observed at low magnetic fields is of the ``uudd"
($\uparrow\uparrow\downarrow\downarrow$)  type along some particular line 
(see for instance  [\onlinecite{Medarde,CheongMostovoy,Khomskii}]
and references therein). 
Such order usually appears when second neighbors antiferromagnetic interactions compete 
with either the uniform or Néel configurations induced by nearest neighbors interactions.
This happens to be the case in quasi-one-dimensional materials like 
Ca$_{3}$CoMnO$_{6}$ [\onlinecite{Cheong-2008}],
quasi-two-dimensional materials like the delafossite 
AgCrS$_{2}$ [\onlinecite{Damay-2011,Streltsov-2015}] 
and also in multiferroic 
manganite perovskites with E-type antiferromagnetic order such as 
HoMnO$_{3}$ [\onlinecite{Dagotto-2006,Dong-2009}], 
ferrite perovskites such as
GdFeO$_{3}$ [\onlinecite{Tokura-2009}]
and other 3D compounds such as the 
CdV$_{2}$O$_{4}$ spinel [\onlinecite{Giovannetti-2011}]
or 
RNiO$_{3}$ nickelates (R=La, Pr, \dots,Lu) [\onlinecite{Catalano-2018}].
Among these $\uparrow\uparrow\downarrow\downarrow$ multiferroic materials, particular interest focuses on double perovskites such as Yb$_{2}$CoMnO$_{6}$ [\onlinecite{Blasco-2017}], 
Lu$_{2}$MnCoO$_{6}$ [\onlinecite{Batista-2011,Chikara-2016}],
Er$_{2}$CoMnO$_{6}$ [\onlinecite{Oh-Oh-2019}], and R$_{2}$NiMnO$_{6}$ (R=Pr, Nd, Sm, Gd, Tb, Dy, Ho, and Er) where a giant magnetoelectric effect has been reported  [\onlinecite{Zhou-2015}].

In a previous paper  [\onlinecite{Cabra-etal-2019}] 
we have introduced a simple microscopic multiferroic model 
describing a system with magnetic and electric dipolar degrees of freedom coupled via lattice distortions. 
This mediated coupling is ubiquitous in magnetoelectric phenomena and may be enhanced 
by the strong influence of the lattice in multilayer multiferroics, as
in some cases the lattice mismatch of the layer and the substrate can generate enormous lattice
distortions and trigger giant multiferroic responses [\onlinecite{Hou-2013,J.White-2017}]. 

In the present work we extend and generalize our previous study in several aspects: 
first and most important, we add antiferromagnetic exchange couplings between next nearest neighbors (NNN) 
reported in most of the above mentioned materials.  When the NNN coupling is strong enough we reproduce 
the experimentally observed $\uparrow\uparrow\downarrow\downarrow$ 
magnetic ordering at zero magnetic field. 
This confirms that magnetic frustration is at the root of the phenomenology observed in many materials.
Second and in order to make closer contact with experiments,  
we introduce an easy axis anisotropy that mimics the effective Ising character observed for otherwise quantum magnetic moments. 
Indeed, the  magnetic ions are inmersed in crystal local fields that generally diminish their quantum character, 
making them 
behave as almost classical Ising variables. 
Good examples of this situation are the spin-ice pyrochlores [\onlinecite{Bramwell-2001}], 
with the exception being Tb based pyrochlores where Ising models seem not to suffice but quantum fluctuations have to be included 
[\onlinecite{Gingras-2014,Rau-2018,Pili-2020}].
Thus a parameter controlling the easy axis anisotropy allows for a phase diagram covering the ``quantum'' and
``classical'' behavior realized in many possible different materials. 
Last but not least, we consider realistic dipolar interactions which
either  from intermediary itinerant electrons [\onlinecite{LiOsO3}], 
from Coulomb forces [\onlinecite{Devonshire-1949}], 
or by other effective mechanism, are expected to act as long range forces. 
Even when truncated at second neighbors, long range dipole-dipole interactions
give rise to new phases in a richer dipole-elastic phase diagram. 

Along this work we discuss the zero temperature ground state of the magnetic, electric and elastic one dimensional system described below.
The main results will be the emergence of a spontaneous bulk polarization at zero magnetic field, 
	as well as a sharp drop thereof once the magnetic field exceeds a critical value.

The paper is organized as follows: 
in Sect.\ \ref{sect:The-model} we define the microscopic model to be discussed, the regions of interest and the methods to be used. 
In Sect.\ \ref{sect: electro-elastic-sector} we explore the behavior of dipolar degrees of freedom in the absence of magnetism, 
finding that long range dipolar interactions give rise to  a new intermediate phase with period three order.
In Sect.\ \ref{sect: spontaneous polarization} we present our main results: 
the spontaneous electric polarization driven by the 
interactions 
and the switch off of this effect as soon as the system is magnetized by an external magnetic field.
In Sect.\ \ref{sect:conclusions} we summarize our results,  
discussing possible experimental tests and applications such as efficient polarization flip devices.

\section{System model and methods\label{sect:The-model}}

\subsection{The model}

The system model under analysis describes magnetic, electric and elastic
degrees of freedom, 
in which magnetic moments and electric dipoles interact independently with the lattice, 
that serves as the intermediary for the effective magnetoelastic coupling we want to describe.

\vspace{3mm}
\paragraph*{Magnetoelastic sector.}
Magnetic ions positions are described as sites $i$ in a linear chain. 
Their regular positions are $x_i=i a $ where $a$ is a lattice constant but 
under distortions the ions move to $x_i+u_i$ along the chain direction, so that sites $i$ and $i+1$ 
will be separated by a distance $a+\delta_i$ with $\delta_i=u_{i+1}-u_{i}$. 
The elastic energy cost of such distortions is given by 
\begin{equation}
  H_\text{elastic}=\frac{K}{2}\sum_{i} \delta_i ^2,
  \label{eq: H elastic}
\end{equation}
where $K$ is the lattice stiffness. 

Magnetic ions themselves are represented by $S=1/2$
spin operators $\mathbf{S}_{i}$ at chain sites.
While the model aims to describe the $\uparrow\uparrow\downarrow\downarrow$ order observed along 
certain lines in two and three dimensional multiferroic materials, 
it is interesting to notice that 
a few compounds that have been identified to become multiferroic 
do show this order in quasi-one-dimensional chains of Cu$^{2+}$ magnetic ions (S = 1/2): 
for instance
LiCuVO$_4$ [\onlinecite{6,7}], 
LiCu$_2$O$_2$ [\onlinecite{8,9,10}], 
CuCl$_2$ [\onlinecite{11}], 
CuBr$_2$ [\onlinecite{12}], 
PbCuSO$_4$(OH)$_2$ [\onlinecite{13,14}], 
CuCrO$_4$ [\onlinecite{15}] 
and SrCuTe$_2$O$_6$ [\onlinecite{SrCuTeO-2016}].

Following our proposal in [\onlinecite{Cabra-etal-2019}], the magnetic ions interact via nearest neighbors
(NN) antiferromagnetic  couplings $J_{1}$.  Frustration is introduced 
by next nearest neighbors (NNN) antiferromagnetic
couplings $J_{2}$.
Both NN and NNN super-exchange couplings have  magnitudes that may depend on elastic distortions. 
However, we assume for simplicity that only the NN exchange shows a linear dependence 
that
can be written as 
\begin{equation}
J_1(\delta_i)=J_1 (1-\alpha \delta_i)\\
\label{eq: J1_i}
\end{equation}
where $\alpha>0$ is called the linear magnetoelastic coupling 
(incidentally, in the frequent case of alternating distortions the 
second neighbor distances $2a+\delta_i+\delta_{i+1}$ are not altered at all). 
Positive $\alpha$ makes NN exchange stronger as magnetic ions approach each other. 

The effect of crystal fields can in general be modeled by anisotropic spin interactions:
the $SU(2)$ invariant Heisenberg interaction $\mathbf{S}_{i}\cdot \mathbf{S}_{j}$ is replaced by 
$S^x_i S^x_j +S^y_i S^y_j+\Delta S^z_iS^z_j$ ($z$ axis determined by the crystal environment).
Aiming to describe collinear multiferroic materials, we focus on $\Delta \geq 1$;
that is, we cover the easy axis anisotropy case $\Delta >1$ 
and in particular the isotropic case $\Delta=1$.
This is motivated by the large variety of known multiferroic materials, 
but also by the theoretical importance of the $SU(2)$ invariant point case.
The easy plane regime $\Delta < 1$,  not discussed here, 
is known to be continuously connected with the isotropic case 
(see for instance [\onlinecite{Giamarchi}]).
On the other hand, the limit $\Delta \to \infty$ connects our work with 
the classical Ising regime.
In order to deal with large $\Delta$ without hiding the other sectors,
we introduce a parameter $\gamma \equiv 1/\Delta$
and absorb $\Delta$ into the exchange constants. 
Finally, we introduce the Zeeman energy associated with an external magnetic field $h$ along the easy axis direction.

The magnetic sector, coupled to lattice distortions,  is then described by the Hamiltonian 
%
\begin{eqnarray}
  H_\text{spin}&=&\sum_{i} J_1(\delta_i) \left(\mathbf{S}_{i}\cdot \mathbf{S}_{i+1}	\right)_\gamma
  +\sum_{i}J_{2}\left(\mathbf{S}_{i}\cdot \mathbf{S}_{i+2}	\right)_\gamma \nonumber\\
  &-&h\sum_i S^z_i,
  \label{eq: H mag}
\end{eqnarray}
where we write for short 
\begin{equation}
\left(\mathbf{S}_{i}\cdot \mathbf{S}_{j}	\right)_\gamma
\equiv 
S_{i}^{z}S_{j}^{z}+\gamma \left(S_{i}^{x}S_{j}^{x}+S_{i}^{y}S_{j}^{y}\right).
\label{anisotropic spin product}
\end{equation}
A model described just by $H_{\text{elastic}}+ H_{\text{spin}}$ might be called a frustrated anisotropic spin-Peierls system.

We recall that the anisotropy parameter $\gamma<1$ ($\Delta>1$)  weakens the
quantum fluctuations of the transverse spin components, making the
spins ``more classical''. 
For systems with collinear order the large $\Delta$ limit is equivalent to considering large $S$ spins,
in the sense that in a Holstein-Primakov [\onlinecite{Auerbach}] expansion transverse fluctuations are suppressed out by a $1/S$ factor. 
Other approaches describe the easy axis component with a strong single ion anisotropy
 [\onlinecite{Seno-1994}], or do instead introduce quantum fluctuations on top of classical spins
  [\onlinecite{Dyson-Maleev-1956,Blanco-2017}].
\begin{figure}[ht]
	\begin{centering}
		\includegraphics[scale=0.7]{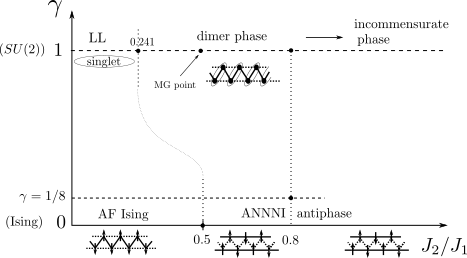}
		\par
	\end{centering}
	\caption{
		Schematic ground state diagram for the spin $S=1/2$ anisotropic frustrated antiferromagnetic chain.
		Most of the materials we are interested in are located in the frustrated anisotropic region
		(low right corner).  
		As representative points we numerically explore in detail a frustration given by  $J_2/J_1=0.8$, 
		in the isotropic case  $\gamma=1$ and a high easy axis anisotropy $\gamma=1/8$. 
	}
	\label{fig: frustration-anisotropy} 
\end{figure}

In the absence of deformations the magnetic model in Eq. (\ref{eq: H mag}) has been thoroughly studied.
We do not intend to cover the subject in all details but summarize the main results relevant 
for the present work;
for a complete treatment with a careful account of the literature see [\onlinecite{Giamarchi}] 
and references therein.
For our purpose it is worth calling to mind its main features
when  no magnetic field is turned on. 
These are governed by the competition between frustration $J_2/J_1$ and anisotropy $\gamma$ and 
can be summarized by the diagram in Fig. \ref{fig: frustration-anisotropy}. 
For low frustration $J_2 \ll J_1$ the system can be seen as a linear antiferromagnetic chain $J_1$ weakly perturbed by NNN interactions $J_2$; 
in the opposite limit $J_2 \gg J_1$ it is better described as two-leg ladder of linear antiferromagnetic chains $J_2$ weakly coupled by zig-zag rungs $J_1$.
The $SU(2)$ symmetric line $\gamma=1$ is well studied by many techniques, 
in particular the bosonization of the effective low energy excitations  [\onlinecite{Haldane-1982}]: 
for low frustration the ground state is a gapless Luttinger Liquid (LL) with quasi long range order
but enters a two-fold degenerate gapped quantum dimer phase for
  $J_2/J_1>0.2411$ [\onlinecite{Okamoto-1992,Eggert-1996}], 
with expectation value of the local spin $\langle S^z_i\rangle =0$ and   
strong  antiferromagnetic (negative) spin correlations every two-bonds
(strictly, this is not collinear).
A paradigmatic example is found at $J_2/J_1=0.5$, the Majumdar-Ghosh point [\onlinecite{Majumdar-Ghosh-1969}], 
where the exact ground state is a (two-fold degenerate) direct product of two-site spin singlets. 
For very large frustration the gap decreases exponentially and the ground state   
shows incommensurate spiral spin correlations [\onlinecite{White-Affleck-1996,Allen-Senechal-1997,Nersesyan-1998}].
On the bottom of the diagram, the large anisotropy limit $\gamma = 0$ defines the one dimensional antiferromagnetic Anisotropic Next Nearest Neighbors Ising (ANNNI) model; 
classical spins order in a two-fold degenerate $\uparrow\downarrow\uparrow\downarrow$  Néel phase for low frustration ($J_{2}/J_{1}<0.5$) 
with a transition to the $\uparrow\uparrow\downarrow\downarrow$ antiphase state for larger frustration ($J_{2}/J_{1}>0.5$)  [\onlinecite{Selke-1988}]. 
In a sense, while $\gamma \to 0$  the LL quantum phase evolves into the classical Néel phase 
and the quantum dimer phase evolves into the $\uparrow\uparrow\downarrow\downarrow$ classical phase. 
Many of the materials we are interested in are located in the frustrated, easy axis anisotropic region 
(low right corner).  
Others correspond to the frustrated  ANNNI model with ferromagnetic $J_1<0$ and antiferromagnetic $J_2>0$, leading to the same  $\uparrow\uparrow\downarrow\downarrow$ antiphase state when $J_{2}/|J_{1}|>0.5$.

\vspace{3mm}
\paragraph*{Electroelastic sector.}
The electric sector is modeled by dipolar moments $\mathbf{p}_{i}$ located midway between magnetic atoms at sites $i$ and $i+1$. 
They might arise from local charge distribution of non-magnetic ions in the crystal unit cell, 
occupying one of two possible Jahn-Teller states determined by the crystal environment
and bridging the super-exchange magnetic couplings. 

As the magnetic ions change their positions, 
the magnitude of dipolar moments may also change. 
It could happen that no local dipolar moment is present in the absence of distortions, 
in this case we would describe the arising dipoles by a magnitude proportional to $\delta_i$ 
and orientation along an appropriate axis. 
For some other materials a local dipolar moment might exist prior to distortions, 
along a given axis $\hat{\mathbf{e}}$.
%
Relatedly, it is worth to recall that the measurable quantity in crystals is not the absolute polarization 
but the polarization change between different states of the same compound [\onlinecite{Resta-1993}]. 
%

\begin{figure}[ht]
\begin{centering}
\includegraphics[scale=0.65]{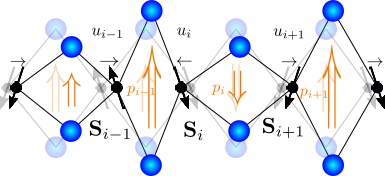} 
\par
\end{centering}
\caption{
Cartoon of degrees of freedom and dipole-elastic coupling mechanism. 
Magnetic ions are described as a chain of sites (in black) while dipoles (double arrows) 
associated to charged ions (in blue) are located at chain bonds. 
Displacements of magnetic ions $u_i$ modify bond lengths by a distortion $\delta_i=u_{i+1}-u_{i}$.
Also the distance between adjacent dipoles is distorted by $\eta_i=(\delta_{i}+\delta_{i+1})/2$. 
A dipole strength is enlarged (shortened) when the bond is shortened (enlarged)  while its orientation is given by an Ising variable, as described in Eq.\ \ref{beta}. 
Shaded symbols show the undistorted (regular) lattice, full colored symbols represent a general distortion configuration.
}
\label{fig: pantograph} 
\end{figure}

\vspace{5mm}
The particular model discussed in this paper is partly inspired in the the material AgCrS$_2$ 
where the magnetic ions Cr$^{3+}$ are surrounded by six S$^{2-}$ non-magnetic sulfur ions on 
the vertices of non regular octahedra (non equivalent crystallographic positions).
It suffers a transition from the paramagnetic  $R3m$ structure
to the $\uparrow\uparrow\downarrow\downarrow$ magnetically ordered phase with non centro-symmetric $Cm$ structure  [\onlinecite{Damay-2011}]. 
This transition produces a magnetostriction enlarging (shortening) the distance between parallel (antiparallel) magnetic moments, 
then producing a shift of the center of charge of surrounding sulfur ions and a consequent spontaneous polarization  [\onlinecite{Streltsov-2015}].
Following this description, and in order to unravel the physical mechanism leading to multiferroicity by lattice distortions, 
we will assume that the undistorted lattice hosts electric dipoles amidst magnetic ions, 
with a natural magnitude $p_0$ and a preferred axis $\hat{\mathbf{e}}$ oriented perpendicular to the chain 
(this choice of axes can be easily generalized to deal with more general situations, but the main novelty of the mechanism presented here is already contained in this simplified description).
Under distortions $\delta_i$ the local dipole magnitude is modified through a pantograph mechanism [\onlinecite{Jaime-2006,Yao-2008,Cabra-etal-2019}]. 
This is modeled in a linear approximation by $\mathbf{p}_i=p_{i}(\sigma_{i},\delta_{i})\hat{\mathbf{e}}$ with a component
\begin{equation}
  p_{i}(\sigma_{i},\delta_{i})=p_{0}\left(1-\beta\delta_{i}\right)2\sigma_{i}.
  \label{beta}
\end{equation}
Here $\sigma_{i}=\pm 1/2$ is an Ising variable for the orientation of the dipole along its axis, $p_{0}$ is the dipolar
moment magnitude in the absence of distortions, and $\beta$ will be called
the dipole-elastic coupling. Notice that $\beta>0$  makes dipolar moments larger as neighboring magnetic sites become closer. 
The pantograph mechanism, depicted in Fig.\ \ref{fig: pantograph},  encodes the interaction between electric dipoles and elastic degrees of freedom. 
Romboids in this picture represent, without loss of generality, the actual crystal environment of magnetic ions.

For a given distribution of distortions $\delta_i$ and dipoles $p_{i}(\sigma_{i},\delta_{i})$ the system acquires a bulk polarization
\begin{equation}
	P \equiv
	 \frac{1}{N_s}\sum_{i=1}^{Ns} p_{i}(\sigma_{i},\delta_{i})
	 = \frac{1}{N_s}\sum_{i=1}^{Ns} p_0 (1-\beta \delta_i) 2\sigma_i,
	 \label{eq: polarization definition}
\end{equation}
where $N_s$ is the chain length (number of sites).

Electric dipolar momenta may interact with each other, at a relevant energy scale, in a phenomenological way. 
Such interaction is  eventually determined by long range dipole-dipole interactions and/or elastic relations 
between deformations of charged and intermediate ions in the crystal [\onlinecite{Spaldin-book}]. 
For the sake of definiteness we consider a Coulomb long range dipole-dipole interaction coupling 
decaying with the cube of the dipole separation,
\begin{equation}
\lambda_D \frac{\mathbf{p}_i \cdot \mathbf{p}_j-3(\mathbf{p}_i \cdot \hat{x})(\mathbf{p}_j \cdot \hat{x})}{|x_j-x_i|^3}
\end{equation}
which in the present geometry only contributes with the product of the transverse components $p_i$. 
Regarding the distance decay,  notice that dipoles $p_{i}$ and $p_{i+1}$ are separated by a distance $a+\eta_i$, 
where $\eta_i=(\delta_{i}+\delta_{i+1})/2$ is the distortion of the distance between adjacent dipoles. 
The electric energy of a given configuration of dipoles coupled to distortions is given by 
\begin{eqnarray}
  H_{\text{dipole}}^\text{(full range)}&=&\lambda_{D}\sum_{i}\left(
  \frac{p_{i}(\sigma_{i},\delta_{i})p_{i+1}(\sigma_{i+1},\delta_{i+1})}{\left(a+\eta_{i} \right)^{3}} \right. \nonumber \\ 
  &+& \frac{p_{i}(\sigma_{i},\delta_{i})p_{i+2}(\sigma_{i+2},\delta_{i+2})}{\left(2a+ \eta_{i} + \eta_{i+1}\right)^{3}} +\cdots\nonumber \\
  &-& E \sum p_{i}(\sigma_{i},\delta_{i}) \label{eq: H dip}
\end{eqnarray}
where the dots represent longer range dipolar interactions and $E$ is an external electric field along the dipolar axis 
$\hat{\mathbf{e}}$. 
An electric field component transverse to this axis would introduce dipolar quantum fluctuations,
interesting in the context of molecular magnets [\onlinecite{Naka-2016}] or the 
ferroelectric SrTiO$_3$  [\onlinecite{SrTiO}] but this is out of the scope of the present work.

We consider here  an expansion of the dipolar interactions in Eq.\ (\ref{eq: H dip}) up to second neighbors.
We expect that the inclusion of longer range terms will not modify qualitatively the arising dipolar phases,
at least for bipartite lattices where further neighbors fall into either the first or the second neighbor sublattices and will only renormalize the frustration.
Assuming small deformations we also expand distortions up to linear terms. We get 
\begin{widetext}
\begin{eqnarray}
H_{\text{dipole}}&=&J_{e}\sum_{i}\left(\sigma_{i}\sigma_{i+1}+\frac{1}{8}\sigma_{i}\sigma_{i+2}\right)
-2\varepsilon\sum_{i}\sigma_{i}
+2\beta\varepsilon\sum_{i}\delta_{i}\sigma_{i} \label{eq: H-dip}   \\
&-&J_{e}\sum_{i}\left[
\left(\beta+\frac{3}{2a}\right)\left(\sigma_{i-1}\sigma_{i}+\sigma_{i}\sigma_{i+1}\right)
+\frac{1}{8}\left(\beta+\frac{3}{4a}\right)\left(\sigma_{i-2}\sigma_{i}+\sigma_{i}\sigma_{i+2}\right)
+\frac{3}{16a}\sigma_{i-1}\sigma_{i+1}
\right]\delta_{i}, \nonumber
\end{eqnarray}
\end{widetext}
where $J_{e}\equiv 4\lambda_D p_0^2/a^3$ and $\epsilon \equiv 2 p_0^2 E$.  A model described by the addition of $H_{\text{elastic}} + H_{\text{dipole}}$ might be called  a dipole-Peierls system.

\vspace{3mm}
\paragraph*{Complete Hamiltonian.}

As we mentioned at the beginning of this Section, 
the elastic degrees of freedom with the Hamiltonian given in  Eq.\ (\ref{eq: H elastic}), 
coupled separately to the spins in Eq.\ (\ref{eq: H mag}) 
and to the dipoles in Eq.\ (\ref{eq: H-dip}), 
are the intermediaries of the magnetoelastic coupling in our proposal. 
This is achieved by the complete Hamiltonian to be discussed below,
\begin{equation}
	H=H_\text{elastic}+H_\text{spin}+ H_\text{dipole},
	\label{eq: H}
\end{equation}
that will be called the spin-dipole-Peierls Hamiltonian.

\subsection{Self-consistent equations}

In order to cope with the three coupled degrees of freedom, one needs an organizing strategy. 
Here we follow a self-consistent method [\onlinecite{Feiguin-etal-1997}]
looking for the elastic distortions that minimize the total energy 
in Eq.\ (\ref{eq: H}).

For a given configuration of dipoles $\sigma_{i}$ and a (quantum or classical) state for the spins $\mathbf{S}_{i}$, 
the minimal elastic energy is obtained when distortions $\delta_{i}$ satisfy the local zero gradient conditions
\begin{widetext}
\begin{eqnarray}
\label{eq: self-consistency}
K\delta_{i}^{\text{free}}&=&\alpha J_{1}\langle S_{i}^{z}S_{i+1}^{z}+\gamma\left(S_{i}^{x}S_{i+1}^{x}+S_{i}^{y}S_{i+1}^{y}\right)\rangle-\beta\varepsilon\sigma_{i}\\
&+&J_{e}\left(\beta+\frac{3}{2a}\right)\left(\sigma_{i-1}\sigma_{i}+\sigma_{i}\sigma_{i+1}\right)
+\frac{1}{8}J_{e}\left(\beta+\frac{3}{4a}\right)\left(\sigma_{i-2}\sigma_{i}+\sigma_{i}\sigma_{i+2}\right)+
J_{e}\frac{3}{16a}\sigma_{i-1}\sigma_{i+1},\nonumber
\end{eqnarray}

\end{widetext}
further constrained by the fixed chain length condition
 $\delta_{i}=\delta_{i}^{\text{free}}-\overline{\delta_{i}^{\text{free}}}$ 
where the bar stands for average value along the chain.

On the one hand these self-consistent (SC) equations clearly exhibit the interplay 
between magnetic and electric degrees of freedom either collaborating or competing to produce the optimal elastic distortions. 
Each of them enters in the form of local correlations.
On the other hand it allows to incorporate the knowledge about the magnetic sector and the electric sector separately. 
It should be stressed that NNN magnetic interactions, although not explicit in Eq. (\ref{eq: self-consistency}),  
play a central role in the actual value of NN correlations 
by introducing magnetic frustration in the Hamiltonian in Eq.\ (\ref{eq: H mag}). 
It is the analysis of this Hamiltonian what allows for theoretical or numerical input into the SC equations. 
Below we both discuss theoretical arguments and provide numerical results by iteratively solving the spin problem with the help of Density Matrix Renormalization Group (DMRG) computations [\onlinecite{White-DMRG}].

We have performed an iterative numerical analysis based on DMRG to solve the magnetic and electric sectors
in the adiabatic equations (6), along the lines stated in [\onlinecite{Feiguin-etal-1997}]
and implemented in a similar context in  [\onlinecite{Cabra-etal-2019}].
The ground state for the spin system is obtained by the DMRG algorithm for each  $\delta_i$ and $\sigma_i$
configuration.  
Therefore, we re-obtain the set of $\delta_i$ from Eq.\ (\ref{eq: self-consistency}) and prove different $\sigma_i$ in order to minimize the total energy.
We have used periodic boundary conditions, 
and we have kept the truncation error
less than O($10^{-12}$), during up to more than 100 sweeps in the worst cases. 
This assures that errors of the DMRG computation are smaller than symbol sizes in each figure.

\subsection{Regions of interest \label{regions of interest}}
 
The various parameters in the model allow for a rich phase diagram. 
According to the multiferroic materials we aim to describe,
the main region of interest along the present work will be that with
large enough ratio $J_{2}/J_{1}$ so as to manifest magnetic frustration. 
For large anisotropy $\gamma \ll 1$ one could expect that spin fluctuations are strongly
diminished, allowing for a ``uudd'' ground state comparable  to the classical ANNNI model antiphase state. 
However, we will show that quantum fluctuations still influence the deep Ising limit.

As for the energy scales, the dipolar exchange $J_{e}$ will be kept below the magnetic exchange couplings, 
so that in principle it is magnetism what drives electric responses. 
The lattice stiffness $K$ will set an energy scale larger than magnetic and electric
ones, in order to keep distortions small with respect to the lattice
spacing $a$.
We set the length scale by taking the lattice spacing $a=1$ and also set the
energy scale taking $Ka^{2}=1$. 

From the above considerations, we choose for numerical computations a reference set of phenomenological parameters 
$J_{1}=0.5$, $J_{2}=0.4$ and $J_{e}=0.2$ to organize the energy scale of each degree of freedom.
We also choose $\alpha=\beta=0.2$ to analyze the magnetoelastic
and electroelastic couplings. Notice that our
results do not depend on fine tuning, so we expect them to be valid
in a wide region of parameters. 

The electric and magnetic fields in Eqs.\ (\ref{eq: H mag}, \ref{eq: H dip}) can be varied in order to set the system
in different polarized and magnetized regimes. Finally, the magnetic anisotropy will
be varied from  the quantum SU(2) symmetric point $\gamma=1$ down to small enough values to explore the large easy axis anisotropy regime 
where classical behavior is expected. 

\section{Polarization process in the presence of an electric field \label{sect: electro-elastic-sector}} 

\begin{figure}[ht]
\begin{centering}
\includegraphics[scale=0.4]{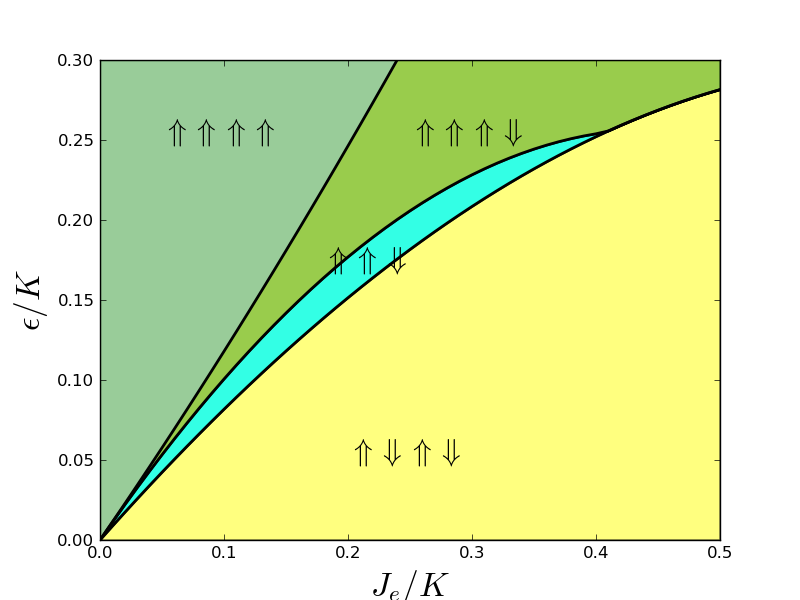} 
\par\end{centering}
\caption{Dipole-elastic phase diagram, computed for $\beta=0.2$. Double line arrows describe the dipole ordered pattern in each region. Elastic distortions follow the dipole pattern periodicity, except in the zero field line $\epsilon=0$ and the saturation region $\Uparrow\Uparrow\Uparrow\Uparrow$ where magnetic ions are equally spaced.  } 
\label{fig: electroelastic} 
\end{figure}

In this Section we discuss the polarization due to an external electric field,
when the magnetic sector is decoupled from the classical degrees of freedom ($\alpha=0$).
%
%
To this end we analyze the minimum energy configurations of the dipole-Peierls Hamiltonian $H_{\text{dipole}}+H_{\text{elastic}}$: 
given different periodic dipolar patterns we analytically compute
the distortions minimizing the elastic energy, in the presence of the electric field.
By comparison we select the lowest energy electroelastic configuration.
In detail, we have considered all of the ordered dipolar configurations up to period four. 
The results lead to the dipole-elastic phase diagram in Fig.\ \ref{fig: electroelastic}. 

It should be stressed that long range dipole-dipole interaction leads to a richer phase diagram,
with respect to the first neighbors interaction case [\onlinecite{Cabra-etal-2019}].  
It includes a new exotic phase where dipoles order with a period of three sites, not found before. 
Distortions occur with the same periodicity  and will eventually contribute or interfere with 
the well-known period three magnetic plateau state that is expected for the 
magnetoelastic sector [\onlinecite{Gazza-2007}]. 
In the present one-dimensional case, and in any bipartite lattice, we expect no other qualitative changes 
by including the interactions between further neighbors.

\begin{figure}[ht]
\begin{centering}
\includegraphics[scale=0.65]{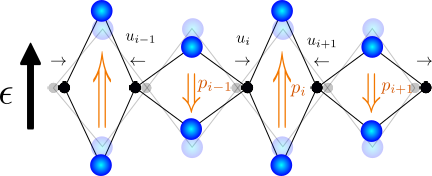}  
\par\end{centering}
\caption{Dipolar pattern in the dimerized electroelastic phase  $\Uparrow\Downarrow\Uparrow\Downarrow$, 
	for a finite electric field pointing upwards (symbols as in Fig.\ \ref{fig: pantograph}). 
	Dipoles pointing along the field are larger than dipoles in the opposite direction. 
	The alternation of bond length distortions is the mechanism for bulk polarization. 
	The system acquires a linear electrical polarization (paraelectric behaviour). 
	Distortions are magnified for visual effect.}
\label{fig: paraelectric pattern} 
\end{figure}

Without electric field the system possesses a $\mathbb{Z}_2$ inversion symmetry, but spontaneously adopts one of the two possible antiferroelectric $\Uparrow\Downarrow\Uparrow\Downarrow$ configurations. To be precise, these are described by
\begin{equation}
  \sigma_i=(-1)^{i+\nu} ,
  \label{eq: sigma AFE}
\end{equation}
where $\nu=0$  ($1$) indicates whether odd (even) dipoles are pointing 
in the positive preferred axis direction. 
The distortions are null in either configuration, then dipoles pointing up or down have the same magnitude and the system  has no net polarization. This is shown with shaded circles in Fig.\ \ref{fig: paraelectric pattern}, with the left-most dipole pointing upwards; the other possibility is got by inversion, or equivalently by a one-site translation.

When a small electric field is turned on, breaking the inversion symmetry, 
no dipole flips are produced below a critical field but dimerized distortions are induced
\begin{equation}
  \delta_i=-(-1)^{i+\nu} \frac{p_0 \beta}{K} \epsilon,
  \label{eq: delta AFE}
\end{equation}
Under these distortions bonds with dipoles pointing along the field get shorter, 
enlarging the corresponding local dipolar momenta 
while bonds with dipoles pointing counter field get longer, 
shortening the corresponding dipolar strength.
This behaviour is sketched in Fig.\ \ref{fig: paraelectric pattern},
and occurs in either antiferroelectric configuration ($\nu=0,\,1$).
The bulk polarization reads
\begin{equation}
  P(\epsilon)=\frac{1}{N_s}\sum_{i=1}^{Ns} p_0 (1-\beta \delta_i) 2\sigma_i= 
  \frac{p_0^2 \beta^2}{K} \epsilon.
\end{equation}
 That is, the system behaves as a simple paraelectric,
 acquiring a bulk polarization proportional to the applied electric field (with electric susceptibility $\chi_e=\frac{\partial P}{\partial E}=\frac{2 p_0^4 \beta^2}{K})$. 

At the critical line that separates the antiferroelectric low field phase from longer period dipolar structures, 
polarization gets discontinuous because of extensive dipolar flips. 
In the present work we concentrate in the low field region properties. 
Discontinuous transitions to higher polarized states, either via an electric or a magnetic field 
and the interplay with magnetization plateaus discussed below will be studied elsewhere. 


\section{Magneto-elastic coupling and spontaneous electric polarization \label{sect: spontaneous polarization}}

When the magnetic sector is coupled to the lattice through $\alpha \neq 0$, 
the ground state magnetic configuration may come along with lattice distortions.
These in turn bring about the possibility
of modulations in the exchange couplings, associated to the lattice distortions. 

In the absence of dipolar degrees of freedom 
this interplay between distortions and modulated exchange couplings is resolved as 
an energy balance between elastic cost and magnetic energy gain. 
Technically, this balance is expressed by self consistent equations similar to our Eqs. (\ref{eq: self-consistency}). 
In general, when non trivial distortions show up in the ground state, the spin excitation spectrum is gapped. 
In consequence the magnetization curve presents a plateau: it requires a finite magnetic field for the Zeeman energy
to overcome the energy gap and change the spin state.
A most important example is the spin-Peierls mechanism that promotes the
formation of spin singlets at the cost of dimerized distortions [\onlinecite{Cross-Fisher-1979,Bray-1982}],
either in the non-frustrated case $J_{2}=0$ or the frustrated one [\onlinecite{Feiguin-etal-1997}]. 
This has been studied not only in one dimensional spin chains but also in higher dimensions
[\onlinecite{Becca-Mila-2002,Klumper-2002,Penc-2004,Klumper-2006,Bissola-2007}].
The spin-lattice coupling also
provides mechanisms for the opening of plateaus at different magnetization
fractions, either for quantum $S=1/2$ spins [\onlinecite{Cabra-Stauffer-2006}]
or classical spins [\onlinecite{Vekua-etal-2006}].

It is important to notice that magnetization plateaus may be related to other mechanisms, 
different from elastic distortions.
One of them is the competition between NN and NNN exchange couplings, 
frustrating the antiferromagnetic order [\onlinecite{Haldane-1982,Okunishi-2003}]. 
Moreover, the easy axis anisotropy drives a competition between the convenience of ground states with quantum
structures (singlets) or classical frustrated configurations [\onlinecite{Hida-Affleck-2005}].

In our system model the magnetic frustration and the magnetoelastic mechanism co-exist, 
along with a dipolar energy cost/gain for lattice distortions. Altogether, this
is expressed in the self consistent Eqs. (\ref{eq: self-consistency}) for lattice distortions.
These SC equations show that the pantograph mechanism puts dipolar and magnetic correlations
in either cooperation or competition with each other to produce changes in the bond lengths.
\textit{This is the key ingredient that provides an effective magnetoelectric coupling mediated by lattice distortions},
opening an avenue to a plethora of new physics.

We show below that this interplay gives rise to a bulk polarization without the presence of an external electric field. 
Moreover, we show that a magnetic field above a threshold causes a sharp  polarization switch. 

To start our analysis we first  address to the existence of a zero magnetization plateau in the magnetization curve 
of the present spin-dipole-Peierls model, at zero electric field. 
As discussed in Section \ref{regions of interest}, 
we focus on the region with high enough frustration so as to produce the $\uparrow\uparrow\downarrow\downarrow$ 
magnetic ordering (see Fig. \ref{fig: frustration-anisotropy});
for numerical work we take as a representative case the parameters $J_1=0.5$. $J_2=0.4$, $J_e=0.2$, $\alpha=\beta=0.2$.
We have  explored the anisotropy range $\gamma \leq 1$  and found signals of quantum and classical behaviour;
we report, as representative examples,  
the $SU(2)$ symmetric case $\gamma=1$ and a highly anisotropic case $\gamma=1/8$.

We solved the self-consistent equations (\ref{eq: self-consistency}) iteratively, 
feeding in the spin-spin correlations computed by DMRG in the presence of distortions 
and the zero electric field antiferroelectric dipolar configuration (see Fig. \ref{fig: electroelastic}). 

\begin{figure}[ht]
\begin{centering}
\includegraphics[scale=0.40]{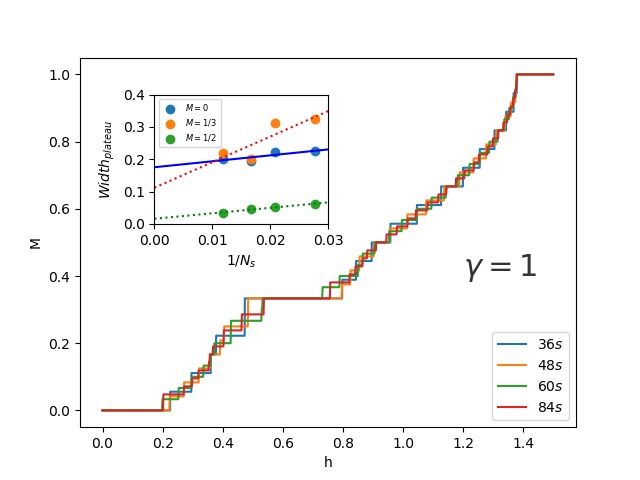} 
\par\end{centering}
\begin{centering}
\includegraphics[scale=0.55]{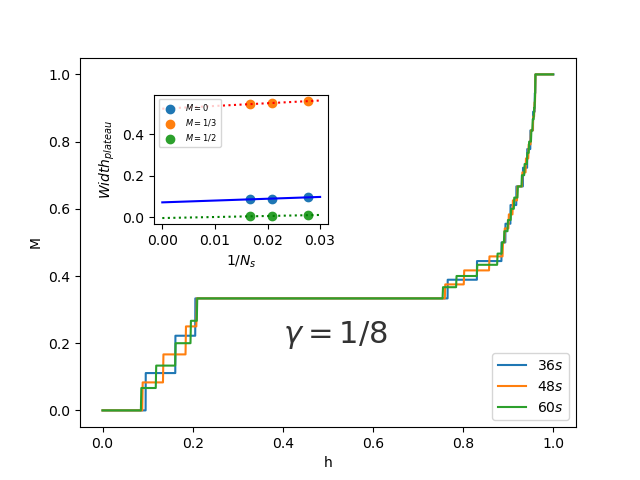}  
\par\end{centering}
\caption{
    Magnetization curves obtained by DMRG self-consistent solution of Eqs.\ (\ref{eq: self-consistency})  for the isotropic case ($\gamma=1$) and  a high easy axis anisotropy ($\gamma=1/8$), setting $J_1=0.5$. $J_2=0.4$, $J_e=0.2$, $\alpha=\beta=0.2$ and zero electric field.
    A plateau at $M=0$ is observed in both cases, though
    the spin structure found for isotropic case (quantum dimerized plateau) is very different from the one found in the anisotropic case (classical $\uparrow\uparrow\downarrow\downarrow$ plateau), see discussion below.
    A prominent plateau at magnetization fraction $M=1/3$ is also observed in both cases. 
    The insets show the finite size scaling of the main plateaus width.
}
\label{fig: magnetization-curves} 
\end{figure}

By covering all the possible magnetizations in a finite size chain of length $N_s$ 
we draw  the magnetization curves shown in Fig.\ \ref{fig: magnetization-curves}
where the magnetization $M$ is defined as the total $\langle S^z_{\text{total}} \rangle$ relative to saturation.

The outcome is a very rich phase diagram that not only includes previously studied situations, 
but also suggests some exotic non-trivial ones.
Besides the $M=0$ plateau present for both the isotropic and the anisotropic case, 
one can see other plateaus at simple fractions of the saturation magnetization.
In particular, there is a noticeable plateau at $M=1/3$ that is much wider in the anisotropic case, 
and comes together with a perid three distortion modulation.
There are also  plateaus at $M=1/2$ and $M=2/3$ in the isotropic case, 
which are no longer present in the anisotropic case. 
In spite of small differences or special points, it is interesting to notice that the bosonization picture, 
strictly valid for $|\gamma| \geq 1$ ($|\Delta| \leq 1$), 
remains essentially the same for the easy axis anisotropic region $\gamma < 1$.

\begin{figure}[ht]
\begin{centering}
\includegraphics[scale=0.7]{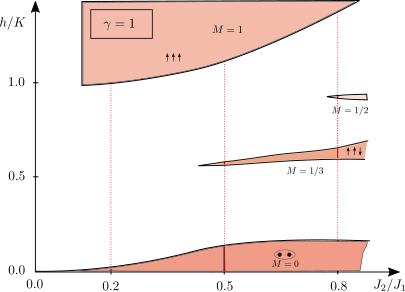}  
\par
\end{centering}
\begin{centering}
\includegraphics[scale=0.7]{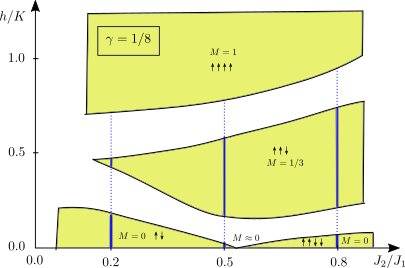}  
\par
\end{centering}
\caption{Schematic magnetic phases showing the appearance of plateaus from
the competition between the frustrated exchange $J_{2}/J_{1}$ and
the magnetic field $h$, as well as the spin-lattice coupling.
We report separately the isotropic case (top panel, $\gamma=1)$ and
a highly anisotropic case (bottom panel, $\gamma=1/8)$. 
$M$ is the magnetization relative to saturation, arrows indicate classical collinear order and points in an ellipse indicate quantum singlet dimers. We do not intend to depict here the   $J_{2}/J_{1}\to 0$ limit.
The vertical lines correspond to magnetization curves at $J_{2}/J_{1}=0.2,\, 0.5 \text{ and } 0.8$, 
where solid segments indicate plateau ranges ($J_1=0.5$, $J_e=0.2$, $\alpha=\beta=0.2$).}
\label{fig: H vs J2/J1} 
\end{figure}

For completeness, we have computed the magnetization curves for systems with some lower frustration values ($J_2/J_1=0.2,\,0.5$).
We sketch in Fig.\ \ref{fig: H vs J2/J1} a 
summary of the observed plateaus in a plane $h$ vs.\ $J_2/J_1$. 

We will focus on the zero magnetization plateau and its magnetic excitations in the present work, 
with emphasis on the description of experimental setups attainable in the multiferroic materials
surveyed in the Introduction. 
Finite magnetization plateaus, which could trigger further experiments in high magnetic fields, 
will be studied elsewhere.

%
%

\subsection{Zero magnetization plateau}

In this Section we compare the magnetic structure of the $M=0$ plateau state observed in the $SU(2)$ isotropic case ($\gamma=1$) 
and the easy axis anisotropic case ($\gamma=1/8$).  
In spite of their differences, we will show that both of them lead to alternating distortions and 
produce a finite bulk polarization  at zero electric field.
Moreover, quantum fluctuations are relevant, though substantially dumped, 
even for the (Ising-like) large anisotropic limit.

\subsubsection{Quantum dimerized plateau}

It is well known that, without exchange modulation ($\alpha=0$)  and with $\gamma=1$, 
the homogeneous isotropic frustrated spin $S=1/2$ antiferromagnetic Heisenberg chain 
spontaneously breaks the translation symmetry and enters a quantum dimer phase for $J_2/J_1>0.2411$ [\onlinecite{Okamoto-1992,Eggert-1996,Giamarchi}], 
with $\langle S^z_i\rangle =0$ and  spin correlations dominated by strong antiferromagnetic (negative) correlations every two-bonds
(strictly, this is not collinear). 
The possibility of forming dimers in even or odd bonds makes the ground state  two-fold degenerate.

In the presence of the magnetoelastic coupling in Eq.\ (\ref{eq: J1_i}) the NN spin-spin correlations 
have influence on elastic distortions, as seen in the first line of Eqs.\ (\ref{eq: self-consistency}).
As the frustrated spin-spin correlations alternate along the chain, 
frustration favors alternating distortions with short bonds accompanying spin singlets. 
Regarding the electroelastic coupling, one can see that the antiferroelectric configuration at zero electric field
has site independent dipole-dipole correlations (negative between first neighbors, positive between second neighbors). 
According to the second line in  Eqs.\ (\ref{eq: self-consistency}), and taking into account the fixed length constraint,  
dipole-dipole correlations have no influence on distortions. 
Thus, our model gets alternating distortions following the frustrated spin correlations. 
The strength of the dipoles sitting in shortened bonds is enlarged, 
while that of dipoles sitting in enlarged bonds is shortened (see Eq.\ (\ref{beta})).
As a consequence  the magnetic frustration gives rise to a \textit{ferrielectric} state, 
carrying a spontaneous bulk electric polarization.
Such a bulk polarization, due to incomplete compensation of local dipole moments, 
has been observed in several multiferroic materials; 
besides the AgCrS$_2$ [\onlinecite{Streltsov-2015}] that motivates our system model, 
well studied materials like
TbMnO$_3$ and TbMn$_2$O$_5$ [\onlinecite{TbMnO3,TbMn2O5}] are clear examples.

Notice that the two-fold degeneracy of the magnetic sector makes it possible 
to locate spin singlets (short bonds) either where dipoles point up or down. 
The spontaneous polarization then has two possible orientations, 
as dictated by the $\mathbb{Z}_2$ inversion symmetry of the model.

The present analysis for the frustrated isotropic magnetoelastic chain 
reinforces our previous results in the absence of frustration  [\onlinecite{Cabra-etal-2019}]
where spontaneous polarization was only due to the spin-Peierls instability of nearest neighbors Heisenberg spin chains.
\begin{figure}[ht]
	\begin{centering}
		\includegraphics[scale=0.4]{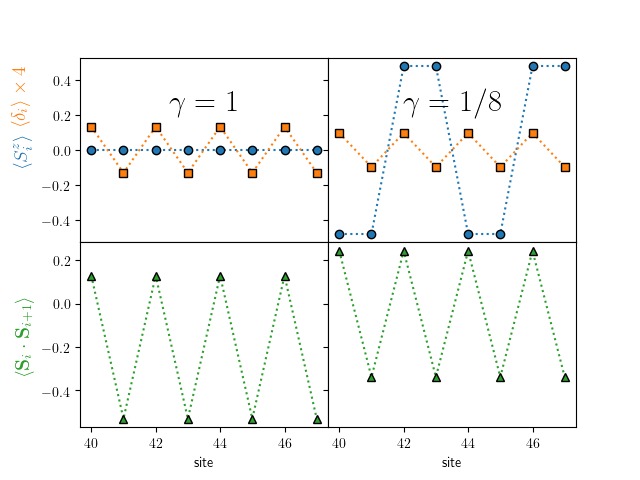}  
		\par\end{centering}
	\caption{Zoomed view of the 
		$M=0$ plateau configuration in a chain of 84 sites 
		with periodic boundary conditions ($J_1=0.5$. $J_2=0.4$, $J_e=0.2$, $\alpha=\beta=0.2$, 
		in the presence of an antiferroelectric dipolar background).
		Upper panels: local profile of $\langle S^z _i\rangle$ (blue circles) and distortions $\delta_i$ (orange squares), 
		in the isotropic case (left panels, $\gamma=1$) and highly anisotropic case (right panels, $\gamma=1/8$). Distortions are scaled by a convenient factor for better visualization.
		Lower panels: local profile of spin correlations $\langle \mathbf{S}_i\cdot  \mathbf{S}_{i+1}\rangle$ in the isotropic and anisotropic cases.
		In the isotropic case the vanishing of $\langle S^z_i\rangle$ and the enhanced alternated
		antiferromagnetic correlations are signals of a quantum dimer phase. 
		In the anisotropic case the consecutive $\langle S^z _i\rangle \approx \pm 0.5$ 
		and  the alternation of ferromagnetic and antiferromagnetic  correlations 
		$\langle \mathbf{S}_i\cdot  \mathbf{S}_{i+1} \rangle \approx \pm 0.25$ 
		indicate a classical  $\uparrow\uparrow\downarrow\downarrow$ phase. 
		In both cases distortions are negative (short bonds) when spin correlations are negative (antiferromagnetic).
	}
	\label{fig: M=0 profiles} 
\end{figure}

\begin{figure}[ht] 
	\begin{centering}
		\includegraphics[scale=0.65]{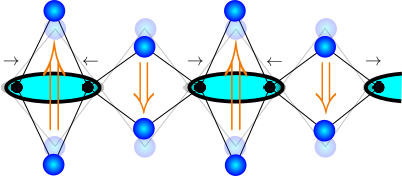} 
		\par\end{centering}
	\caption{ Schematic picture for the quantum plateau state at $M=0$. 
		The two-spin singlets represented by ellipses gain magnetic energy by shortening their distance, thus enlarging their exchange coupling. The influence of these distortions on the alternating dipoles lengths (double arrows) produces a ferrielectric configuration with a finite bulk polarization.
	}
	\label{fig: polarized quantum dimers} 
\end{figure}

For concreteness, we show in the left panels of Fig.\ \ref{fig: M=0 profiles} the local spin expectation value,  the distortion profile and spin-spin correlations obtained by solving 
Eqs.\ (\ref{eq: self-consistency})  for $J_1=0.5$. $J_2=0.4$, $J_e=0.2$, $\alpha=\beta=0.2$, $\gamma=1$.
As anticipated, there are alternating lattice distortions. 
The local magnetization vanishes, $\langle S^z_i\rangle=0$. 
The spin-spin correlations are strongly antiferromagnetic where the  bonds are shortened, 
and weakly ferromagnetic along enlarged bonds; 
this indicates the formation of spin singlets every two bonds and defines the quantum dimer phase. 
A second degenerate solution looks the same, but with dimers translated by one lattice site. 
A pictorial description of this states, including enlarged  dipolar moments at singlet bonds, is shown in Fig. \ref{fig: polarized quantum dimers}.

In the presence of an electric field (not enough to produce dipole flips, see Fig. \ref{fig: electroelastic}), 
the dipole-field term in the SC equations also favors the alternation of distortions. 
But now it selects the short bonds to be located where dipoles point along the field 
(as already discussed in the electroelastic sector, see Section \ref{sect: electro-elastic-sector}). 
In other words, an infinitesimal poling electric field breaking the $\mathbb{Z}_2$ symmetry 
is enough to select one of the otherwise degenerate electric polarization states of the system.

\subsubsection{Classical $\uparrow\uparrow\downarrow\downarrow$ plateau}

In the easy axis anisotropy limit $\gamma \to 0$ and no magnetoelastic coupling  ($\alpha=0$) 
our model coincides with  the homogeneous frustrated antiferromagnetic Ising chain (ANNNI model). 
It is known that this model enters the collinear antiphase 
($\uparrow\uparrow\downarrow\downarrow$) state at $J_2/J_1>0.5$ [\onlinecite{Selke-1988}],
where $J_2$ is large enough to make the NNN spin correlations everywhere antiferromagnetic, 
	while NN correlations alternate between values $\pm S^2$.
Same as in the quantum case, the analysis of the self-consistent conditions in Eq.\ \ref{eq: self-consistency} 
shows that the magnetoelastic terms favor alternating distortions, inducing the $\mathbb{Z}_2$-symmetric
spontaneous polarization.

\begin{figure}[ht]
	\begin{centering}
		\includegraphics[scale=0.65]{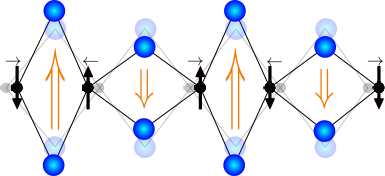} 
		\par\end{centering}
	\caption{Schematic picture for the $\uparrow\uparrow\downarrow\downarrow$ plateau state at $M=0$. 
		The collinear spin configuration  represented by black arrows gains magnetic energy by enlarging 
		the  exchange coupling of anti-parallel nearest neighbors, shortening their distance. Same as in the quantum dimerized plateau, the influence of these distortions on the dipole strengths (double arrows) produces a ferrielectric configuration with a finite bulk polarization.
	}
	\label{fig: polarized uudd} 
\end{figure}

To explore this classical scenario we performed the self-consistent DMRG computation of the ground state for the same parameters as in the previous subsection, 
but for a markedly anisotropic easy axis spin-spin interaction, $\gamma=1/8$. We show in the right panels of Fig.\ \ref{fig: M=0 profiles} the spin and distortion profiles.
They indicate that the spins almost saturate the $z$ component, $\langle S^z_i \rangle \approx \pm 1/2$, following the $\uparrow\uparrow\downarrow\downarrow$ pattern. 
Spin-spin correlations are close to classical, with  $\langle \mathbf{S}_i\cdot  \mathbf{S}_{i+1}\rangle\approx 1/4$ for ferromagnetic bonds and $-1/4$ for antiferromagnetic bonds. 
The distortions do alternate, with short (long)  bonds when spin correlations are antiferromagnetic (ferromagnetic). A graphical description of this state is shown in Fig.\ \ref{fig: polarized uudd}.
Same as in the quantum dimerized plateau, alternating distortions lead to a finite spontaneous electric polarization.

It is worth emphasizing the robustness of the spontaneous polarization induced by magnetic instabilities in the pantograph model.
We have found the same result in very different regimes, 
such as the magnetically frustrated $J_1-J_2$  quantum spin chain, the close to classical frustrated (Ising) chain, 
and the spin-Peierls chain without magnetic frustration [\onlinecite{Cabra-etal-2019}]. 
However, an easy axis anisotropy in the low frustration regime 
(lower left region in Fig.\ \ref{fig: frustration-anisotropy}) induces a Néel antiferromagnetic state 
with homogeneous correlations [\onlinecite{Inagaki-1983,Li-Chen-2010}] that would eventually destroy the spontaneous polarization. 

\subsection{Magnetic excitations}

The $M=0$ configuration remains stable under an external magnetic field $h$, 
until it reaches a critical value $h_c$ such that the gain in Zeeman energy of a magnetically excited state is larger than the spin gap. 
In this situation the system overpasses the $M=0$ plateau and enters a magnetized regime (see Fig.\ \ref{fig: magnetization-curves}). 
In order to understand the magnetization process we start by analyzing the features of the $S^z_\text{total}=1$ state; 
we then check that low magnetization states can be described as a superposition of elementary magnetic excitations. 

\subsubsection{Excitation of the quantum dimerized plateau}

There exist extensive studies of the  $S^z_\text{total}=1$  excitation of the  $S=1/2$ magnetoelastic spin-Peierls Heisenberg chain,
which appears to be fractionalized into two spinons [\onlinecite{Arai-1996}]. 
In the bosonization framework 
these spinons can be explained as topological solitonic excitations 
of a sine-Gordon low energy effective continuum theory coupled to the distortion field [\onlinecite{Fukuyama-1980}].
Their presence has been checked numerically by different techniques [\onlinecite{Feiguin-etal-1997}]
and they are found to condense at the ground state in the presence of a magnetic field.

\begin{figure}[ht]
	\begin{centering}
\includegraphics[scale=0.6]{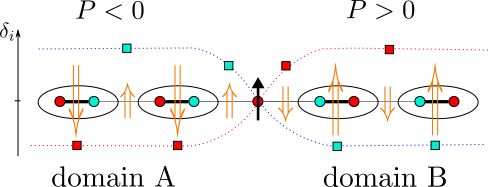} 
		\par\end{centering}
	\caption{
		A magnetic soliton connects the two possible quantum dimer vacua.  In this qualitative picture red (cyan) circles represent odd (even) magnetic sites; 
		squares represent the distortions of the bonds at the right of sites with the same color, dotted lines are a guide to follow odd and even site distortions; double arrows represent electric dipoles sitting amidst magnetic sites, in an antiferroelectric configuration.  
		The sequence of spin singlets (ellipses, thick lines indicating enhanced NN exchange) is shifted by one lattice site across the soliton, 
		defining a different dimerized domain. A spin $S=1/2$ (black arrow) indicates the fractional magnetization carried by the soliton. 
		The sequence of short-long bonds is shifted accordingly. 
		In the presence of the antiferroelectric dipolar background the dimerization
		defines ferrielectric domains with opposite polarization.  
	}
	\label{fig: site-shift} 
\end{figure}

Relevant to our purpose is the fact that the topological solitons connect different degenerate vacua of the system. 
In the spin-Peierls Heisenberg chain the ground state is two-fold degenerate and these vacua are the two possibilities 
of forming singlet pairs along the chain; that is, the two vacua differ by a one-site translation. 
The sequence of elastic distortions is also shifted by one site across each soliton, 
as the short bonds belong together with magnetic singlet pairs.
We call each of these vacua a dimerized domain, say A and B.
A qualitative picture in Fig.\ \ref{fig: site-shift} illustrates the 
two different dimerized domains separated by a soliton.

We have checked numerically that solitons also develop in the present model,
when distortions are coupled to the amplitudes of antiferroelectrically ordered dipoles. 
As well, the distortion pattern shows two different domains A and B, separated by the magnetic solitons.
\textit{At each domain the dipoles develop a ferrielectric net polarization, pointing in opposite directions}.
It is important that both domains are found to have approximately the same length. 
This is expected from the sine-Gordon low energy theory [\onlinecite{Manton_book}] 
and  numerically observed [\onlinecite{Mastrogiuseppe-2008}] due to the exponential tails of the soliton profiles, 
which produce a residual repulsion between them. 
It has been shown that for higher $S^z_\text{total}$ the excitations are pairs of solitons distributed as a periodic array, 
evolving into a sinusoidal magnetization profile [\onlinecite{{Lorenz_1998}}].

\begin{figure}[ht]
	\begin{centering}
		\includegraphics[scale=0.4]{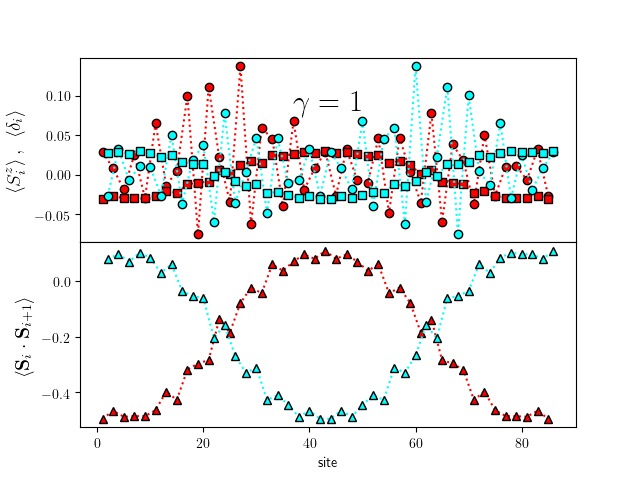} 
		\par\end{centering}
	\caption{
		Local observables in the
		$S^z_\text{total}=1$ excitation 
		in a chain of 86 sites, in the isotropic case $\gamma=1$ 
		(with periodic boundary conditions, $J_1=0.5$, $J_2=0.4$, $J_e=0.2$, $\alpha=\beta=0.2$,
	in the presence of an antiferroelectric dipolar background). Notice that, using periodic boundary conditions, 
	we have changed the chain length to 86  sites 
	for commensurability of the DMRG solution of the $S^z_\text{total}=1$ excitation with the lattice size.
		Upper panel: local profile of $\langle S^z _i\rangle$ (circles) and distortions $\delta_i$ (squares).  
		Lower panel: local spin correlations $\langle \mathbf{S}_i\cdot  \mathbf{S}_{i+1}\rangle$ (triangles). 
		The same colors red, cyan in Fig. \ref{fig: site-shift} 
		are used here to visually distinguish odd, even sites and bonds.		
		The magnetic state develops a two-soliton profile for spin correlations, 
		separating equal length  domains. The dimerized distortions follow the same profile.
		The local magnetizations $\langle S^z _i\rangle$ do not vanish, being larger around the soliton regions, but show no clear order. 
		Qualitative features agree with the cartoon in Fig. \ref{fig: site-shift}.
		}
	\label{fig: quantum excited profiles} 
\end{figure}

Our numerical results for the $S^z_\text{total}=1$ excitation are shown in  Fig.\ \ref{fig: quantum excited profiles}.
Detailed data shows that the distortions (squares in the upper panel) in odd/even sites 
are interchanged across the first soliton, as sketched with the same colors in Fig.\  \ref{fig: site-shift},
and interchanged again to its original sequence across the second soliton,  
so that the short/long bond sequence is shifted by one site at each soliton.
The alternation of ferromagnetic/antiferromagnetic correlations (triangles in the lower panel)  
follows the same sequence as distortions, indicating singlets in two different dimerized domains.  
The magnetic excitation is localized in the soliton regions, 
with an incipient  $\langle S^z_i\rangle$ spin component.
As the soliton regions are wide,  in the finite lengths accessible by DMRG 
	$\langle S^z_i\rangle$ does not reach the null value seen in the vacuum state; 
	instead, distortions and correlations clearly reach their vacuum patterns.
Similar behavior has been reported for the excitation of fractional plateau states in the frustrated magnetoelastic spin chain 
[\onlinecite{Gazza-2007,Hida-Affleck-2005}]. 

We have also studied higher magnetically excited states, where a  pattern of equidistant soliton pairs shows up, 
thus confirming the described cancellation of electric polarization.

\subsubsection{Excitation of the $\uparrow\uparrow\downarrow\downarrow$ plateau}

Given the Ising-like $\uparrow\uparrow\downarrow\downarrow$ structure found 
in the anisotropic case $\gamma=1/8$
for the $M=0$ plateau 
in Fig.\ \ref{fig: M=0 profiles} (right panels), 
one could expect that the $S^z_\text{total} =1$ magnetic excitation   also looks Ising-like, 
that is a simple spin flip followed by a rearrangement of classical spins defining sharp domain walls where some
second neighbors correlations get frustrated (ferromagnetic).
%
\begin{figure}[ht]
	\begin{centering}
		\includegraphics[scale=0.4]{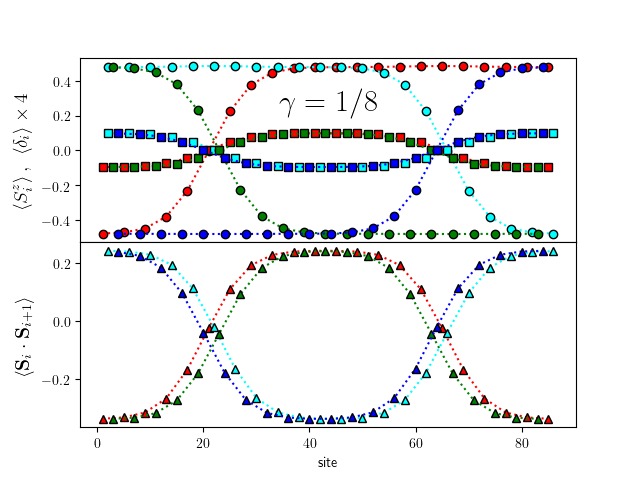} 
		\par\end{centering}
	\caption{
		$S^z_\text{total}=1$ excitation configuration in a chain of 86 sites 
		with periodic boundary conditions ($J_1=0.5$, $J_2=0.4$, $J_e=0.2$, $\alpha=\beta=0.2$,
		in the presence of an antiferroelectric dipolar background), 
		in the anisotropic case $\gamma=1/8$.
		Upper panels: local profile of $\langle S^z _i\rangle$ (circles) and distortions $\delta_i$ (squares), 
		in the isotropic case ($\gamma=1$) and highly anisotropic case ($\gamma=1/8$). 
		Lower panels: local profile of spin correlations $\langle \mathbf{S}_i\cdot  \mathbf{S}_{i+1}\rangle$ (triangles). Four different colors red, cyan, green, blue are used to visually help the location of data every four sites.		
		The system develops two magnetic solitons, separating equal length domains.
		The $\uparrow\uparrow\downarrow\downarrow$ spin pattern, as well as the short/long distortion pattern, 
		are shifted by one lattice site across each soliton.
	}
	\label{fig: ANNNI excited profiles} 
\end{figure}

However, it happens that the system takes advantage of quantum fluctuations to develop solitonic excitations, 
so that the reduction of $\langle S^z_i\rangle$ in the soliton region lowers the energy cost
of the frustrated second neighbors correlations.
Away from the soliton regions, the same as in the quantum case, 
we find that the alternation of distortions and spin correlations, 
and the saturated $\uparrow\uparrow\downarrow\downarrow$ spin pattern, 
are similar to the classical $S^z_\text{total}=0$ plateau structure but shifted 
by one lattice site across each soliton. 
The same as in the isoptropic case, \textit{the electric polarization forms ferrielectric domains with the polarization
pointing in opposite directions}.

\begin{figure}[ht]
	\begin{centering}
		\includegraphics[scale=0.55]{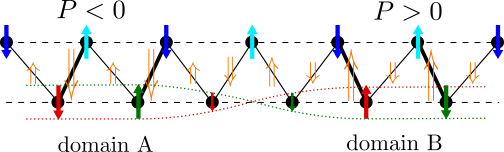}
		\par\end{centering}
	\caption{
		Schematic description of the first soliton in Fig.\ \ref{fig: ANNNI excited profiles},
		connecting two different $\uparrow\uparrow\downarrow\downarrow$ dimerized domains. 
		The linear spin chain can be followed along the rungs of a zig-zag ladder, separating odd sites in the lower leg and even sites in the upper leg. 
		Nearest neighbor exchanges $J_1$ are represented by solid  rung lines 
		and next nearest neighbor exchanges $J_2$ by dashed straight leg lines.
		Single arrows represent the $\langle S^z_i \rangle$ component of spins, using the same sublattice colors as in Fig.\ \ref{fig: ANNNI excited profiles}, 
		and orange double arrows represent the electric dipoles between them.
		Notice that the $\uparrow\uparrow\downarrow\downarrow$ spin order along the chain can be seen as Néel configurations along each leg.
		The magnetic soliton reverses the spins in the lower leg (indicated as a twist in the dotted lines), 
		leaving unchanged those in the upper leg. 
		The lattice dimerization brings closer the antiparallel nearest neighbor spins, enlarging their exchange couplings (thick solid rungs); 
		thus the magnetic twist produces different dimerization domains with ferrielectric polarizations in opposite directions.	
	}
	\label{fig: cartoon uudd} 
\end{figure}
We  show in Fig.\ \ref{fig: ANNNI excited profiles} these results for anisotropy $\gamma=1/8$ 
(\textit{cf.} the $M=0$ state in Fig.\ \ref{fig: M=0 profiles}, right panels), 
using a sequence of colors to identify four sublattices. 
Spins at the left side show a vacuum $\uparrow\uparrow\downarrow\downarrow$ configuration; spins at odd sites 
(red and green sublattices) are flipped across the first soliton to connect with a
different $\uparrow\uparrow\downarrow\downarrow$ vacuum; the same happens with  spins at even sites 
(cyan and blue sublattices) across the second soliton. 
Distortions are dimerized, changing the dimerization domain 
across each soliton. 
Spin correlations in the lower panel show that antiparallel spins go along with short bonds in vacuum regions, 
but quantum fluctuations fade away the expectation value and correlations of spins in the soliton regions. 
This fact reduces the energy cost of the solitonic ``domain wall'', as compared with sharp classical domain walls. 
As a visual aid, we summarize in Fig.\ \ref{fig: cartoon uudd} the $\uparrow\uparrow\downarrow\downarrow$
soliton features in a cartoon picture.

Notice that the solitons in the anisotropic case are slightly narrower than those in Fig.\ \ref{fig: quantum excited profiles}, 
for the isotropic case $\gamma=1$. 
The more anisotropic the interaction, we have checked numerically that the soliton regions gets even narrower. 
But they do not evolve into sharp domain walls, 
at least for anisotropies as large as $\gamma = 0.01$ ($\Delta = 100$).  
It is remarkable that quantum fluctuations play a significant role even in the quasi-classical limit.

The presence of topological solitons, instead of sharp domain walls, is decisive in the formation of equal length 
$\uparrow\uparrow\downarrow\downarrow$ domains: 
it is the repulsive residual interaction between solitons what keeps them separated in the finite size chain.

\subsection{Polarization jump driven by magnetic field}

At zero electric field, both in the isotropic and the anisotropic cases, 
the solitonic magnetic excitations separate ferrielectric domains with opposite polarization.  
This happens not only for $S^z_\text{total}=1$ but for higher excitations described by pairs of solitons.
As a consequence, having these domains the same length, the total polarization of the system drops nearly to zero. 
That is, \textit{the spontaneous electric polarization observed at zero magnetization is switched off 
by means of the applied magnetic field} [\onlinecite{Cabra-etal-2019}]. 
This happens either if the exit from the $M=0$ plateau is smooth 
(that is, soliton pairs appear continuously with the magnetic field) 
or in the case of a metamagnetic jump in which soliton pairs proliferate.

\begin{figure}[ht]
\begin{centering}
	\includegraphics[scale=0.5]{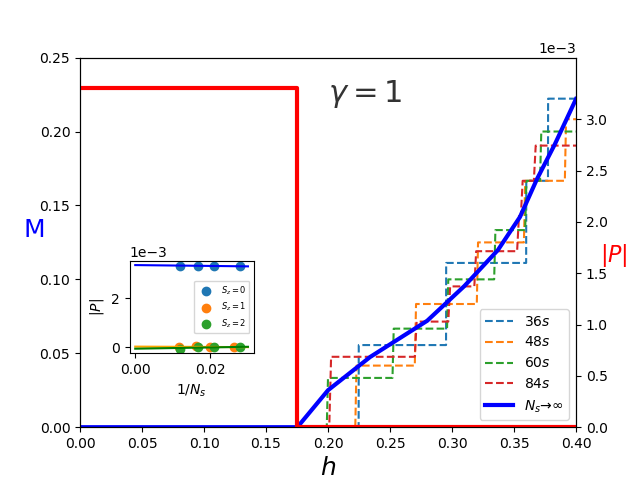}
	\includegraphics[scale=0.5]{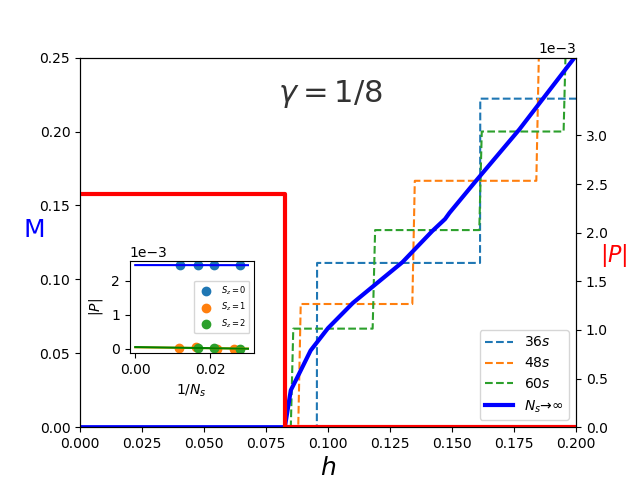} 
\par\end{centering}
\caption{
	Polarization curves (red solid lines, scale in the right axis in units of $p_0$) in an external magnetic field 
	for the isotropic $\gamma=1$ and the anisotropic $\gamma=1/8$ models
	($J_1=0.5$, $J_2=0.4$, $J_e=0.2$, $\alpha=\beta=0.2$,
	in the presence of an antiferroelectric dipolar background).
	Magnetization curves in the low $M$ region
	(extrapolated as blue solid lines, scale in the left axis)
	are also plotted for comparison.
	In both cases the system supports a finite spontaneous polarization at low magnetic fields, while $S^z_\text{total}=0$,
	but a sudden drop is observed once the system exits the $M=0$ magnetic plateau. 
The polarization curves follow from finite size results and infinite size extrapolation. 
Insets: finite size scaling for the polarizations obtained for $S^z_\text{total}=0,\, 1,\, 2$ shows almost no size dependence. }
\label{fig: polarization switch} 
\end{figure}

To make apparent the relation between the polarization jump and the onset of magnetization, 
we plot together in Fig.\ \ref{fig: polarization switch} 
the polarization and the low magnetization curves in a magnetic field, both for the isotropic (upper panel) 
and the anisotropic (lower panel) cases discussed along this work.
The spontaneous polarization (red curves, scale in right axis) is computed from the lattice distortions in an antiferroelectric background, 
according to Eq.\ \ref{eq: polarization definition}. In both cases it suddenly drops several orders of magnitude.
The magnetization is the same as in Fig.\ \ref{fig: magnetization-curves}, with the addition of an infinite size extrapolation (blue curves, scale in left axis).
The infinite size extrapolation of the polarization at the lowest magnetization levels, shown in the insets, 
clearly proves that the polarization switch off is a bulk magnetoelectric effect occurring at the onset of magnetization. 
Beyond the excited $S^z_\text{total}=1$ and $S^z_\text{total}=2$ states, with polarization shown in the insets, 
we have checked that the further increase of the magnetization introduces extra pairs of solitons. 
These appear uniformly spread along the chain, as it also occurs in the magnetoelastic case [\onlinecite{Lorenz_1998}],
separating different dimerization domains and producing the drop of the electric polarization observed 
in Fig.\ \ref{fig: polarization switch} for arbitrary non vanishing magnetization.

Such magnetically driven polarization jumps are a source of intrigue in many multiferroic materials. 
For instance, Lu$_2$MnCoO$_6$  [\onlinecite{Chikara-2016}] 
and  Er$_2$CoMnO$_6$ [\onlinecite{Oh-Oh-2019}] show a polarization jump when exiting 
the observed $M=0$ magnetization plateau. 
Closely related are the polarization jumps observed in R$_2$V$_2$O$_7$ (R = Ni, Co)
when entering and exiting 
the $M=1/2$ magnetization plateau [\onlinecite{Ouyang-2018}]. 
We expect that the present results could help in fitting actual parameters in these materials and explain the observed jumps.

\subsection{Polarization flip controlled by very low electric fields \label{sect: device}	}

Measures of spontaneous polarization are usually made with the help of a tiny poling field, 
to lift the degeneracy between the possible spontaneous orientations. 
Once done, a coercive field much larger than the poling one is required to flip the bulk polarization.

In the present model it is also interesting to discuss the effect of a poling electric field
when the polarization has been switched off by a magnetic field larger than the critical one, 
strong enough to magnetize  and depolarize the system 
by the creation of pairs of different ferrielectric domains.
It happens that the domains with polarization pointing along the poling field are energetically favored, 
hence pushing apart the soliton walls at their ends to lower the total system energy.
As the displacement of solitonic domain boundaries in large systems has very small energy cost, a high electric susceptibility 
is expected in this regime. 
In consequence, the polarization cancellation is not perfect and the system exhibits a 
small net  polarization in the direction of the electric field. 
From this situation, as soon as the magnetic field is turned off, 
it is expected that the orientation of the much larger recovered spontaneous polarization
follows the preferred orientation set by the poling field in the magnetized regime. 
\begin{figure}
	\begin{centering}
		\includegraphics[scale=0.75]{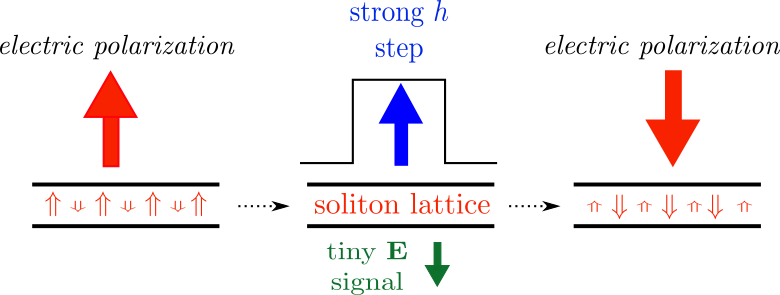}
		\par
	\end{centering}
	\caption{A magnetic field step, in any orientation and strong enough to magnetize the system, produces electric depolarization.
		In combination with a tiny poling field signal, it can be used to reverse the spontaneous polarization $\mathbf P$. 
		This could be the basis for storing information in a dipolar memory bit.
	}
	\label{fig: device} 
\end{figure}

One can think of designing a multiferroic memory storage in which information, in the form of a polarized spot, 
is controlled by a low electric field signal with the help of a brief but strong magnetic blast: 
a magnetic field, carrying no information, would erase the previously "written" polarization, 
which is then "rewritten" in the desired (up or down) orientation 
by the simultaneous presence of a poling low electric field (low voltage bias). 
The procedure is sketched in Fig.\ \ref{fig: device}.
Such a device would show a giant electric response, and could be the basis for an efficient memory writing/reading device.

\begin{figure}[ht]
\begin{centering}
	\includegraphics[scale=0.35]{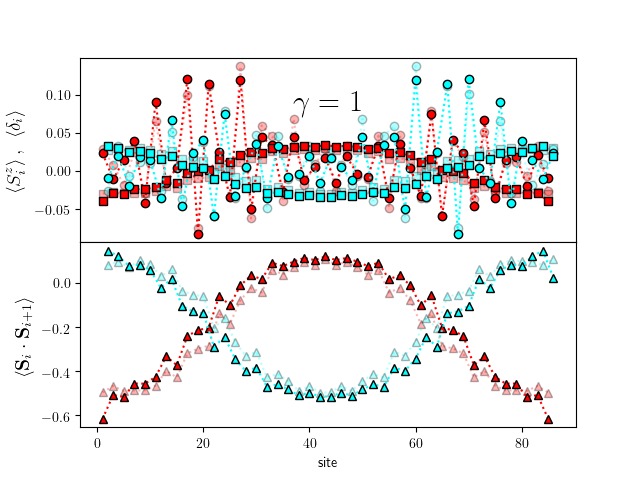}  
	\includegraphics[scale=0.35]{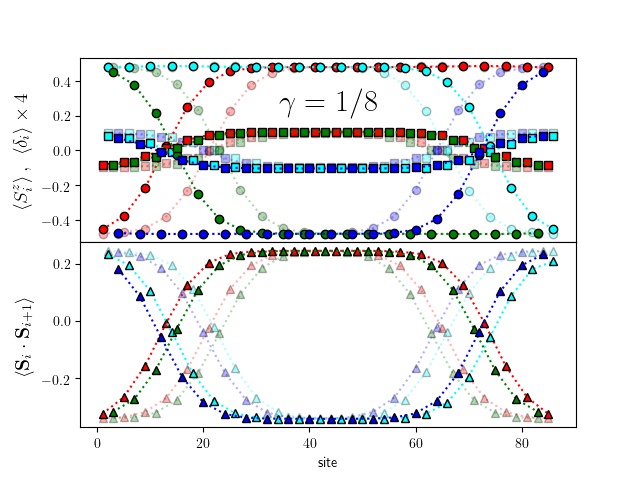}  
	\par\end{centering}
\caption{A small electric field provokes the displacement of the solitons in the $S_z=1$ configuration, 
	enlarging the domain with electric polarization along the field. 
	The solid symbols correspond to an electric field $\epsilon=0.01$, with the rest of the parameters as in Figs.\ \ref{fig: quantum excited profiles}, 
	\ref{fig: ANNNI excited profiles} (repeated here in shaded symbols for comparison) .
	The same effect is observed both in the isotropic and the anisotropic case. 
}
\label{fig: small E} 
\end{figure}

In order to support these considerations we show in Fig.\ \ref{fig: small E} how the equal length domains already seen in Figs.\ \ref{fig: quantum excited profiles}
and \ref{fig: ANNNI excited profiles} (shaded symbols here)
are modified in the presence of a small electric field $\epsilon=0.01$: the central domain, 
with polarization along the field, indeed gets wider.  
We note that in the computationally accessible finite size chains the effect is more pronounced in the anisotropic case, 
where the solitons are narrower and their residual repulsion is less manifest.

\section{Summary and perspectives \label{sect:conclusions}}

In the present work we have extended and improved a microscopic mechanism of magnetoelectric coupling mediated by lattice distortions,
previously introduced by the authors in [\onlinecite{Cabra-etal-2019}], into a realistic model for type II collinear multiferroic materials. 
Essential ingredients to match with experimental observations are the easy axis anisotropy $\Delta > 1$ 
(expressed for convenience as  $\gamma <1$) favoring collinearity, the magnetic frustration $J_2/J_1$ leading to the ``uudd'' spin ground state  
and the Coulomb-like long range dipole-dipole interaction establishing the antiferroelectric order, 
all of these in the absence of external fields. 
Motivated by the variety of known multiferroic materials,
which includes the $SU(2)$ symmetric as well as strongly easy axis anisotropic  spin interactions, 
we have explored the proposed model from the Heisenberg isotropic regime $\Delta=1$  up to  Ising-like anisotropic cases $\Delta \gg 1$. 

The microscopic mechanism may be described by a spin-dipole-Peierls Hamiltonian, 
where the indirect magnetoelectric coupling  arises from a combination of a spin-Peierls like magnetoelectric coupling,
which is known to lead to an elastic dimerization instability, 
and a pantograph mechanism that relates the strength of electric dipolar moments to lattice deformations.
Both mechanisms are ubiquitous in multiferroic materials, 
specially when competing magnetic interactions frustrate an antiferromagnetic Néel configuration.
Magnetic and electric degrees of freedom can thus either cooperate or compete in provoking lattice instabilities,
in a precise way expressed in the selfconsistent key Eqs.\ \ref{eq: self-consistency}.

We have argued theoretically and proven numerically, by extensive DMRG computations, 
that in a wide parameter region, starting at the isotropic $SU(2)$ Heisenberg model 
and going up to an extreme anisotropic ANNNI model, 
the system has a gapped magnetic ground state associated to dimerized lattice distortions.
Main consequences are the zero magnetization plateaus in the magnetization curves and the emergence of
an spontaneous ferrielectric bulk polarization (an antiferroelectric with a remanent polarization), 
with two possible degenerate orientations ($\mathbb{Z}_2$ symmetry).   

In the presence of an external magnetic field exceeding a critical value,
related to the spin gap,  low 
magnetization excitations develop as pairs of topological solitons 
that separate  different dimerized domains carrying opposite ferrielectric polarizations. 
A lattice of equidistant solitons grows along the system, 
producing a sharp switch off in the bulk polarization.
This mechanism, robust due to its topological character, 
could be at the root of the  bulk polarization jumps observed in many different
multiferroic materials. 
We expect that the present paradigm might be fitted to actual experimental parameters and 
be identified as one of the microscopic mechanisms behind magnetically induced polarization jumps.

We have also found a novel polarization state at intermediate electric fields 
with $\Uparrow\Uparrow\Downarrow$ periodicity,
exclusively due to the long range character of the dipolar interacions 
frustrating the antiferroelectric order. 
Such a period three dipolar configuration, 
combined with the $M=1/3$ magnetic plateau state found at intermediate magnetic fields,
could give rise to interesting magnetoelectric cross effects. This will be studied elsewhere.

Regarding technological interest, a material described by our model is classified as a \textit{ferrielectric}. 
It has a spontaneous $\mathbb{Z}_2$ polarization due to dipolar imbalance that can be easily controlled by applied fields.
In fact the presence of a small poling electric field 
gives rise to a relative displacement of the solitonic domain walls,
making the polarization of the magnetized states not to be completely turned off. 
Then a demagnetization would select a preferred orientation for the spontaneous polarization.
This property could be used, for instance, to engineer polarized memory storage devices controllable by very low electric signals. 
From a different point of view, the present work could guide the design and manufacture of composite artificial multiferroic systems, 
such as  multilayers (see for instance [\onlinecite{Taniyama-2015}]) 
where the mechanical strain transfer couples ferroelectricity and ferromagnetism, 
or even regularly nano-patterned arrays (see for instance [\onlinecite{Cai-2018}]) 
where flexoelectricity couples magnetostrictive strain gradients with electric polarization, in different materials.
The technological control of multiferroicity in these multiphase composite systems is rapidly progressing 
and could in a future be the alternative to chemically synthesized multiferroic compounds.
We hope that the understanding of the mechanisms of multiferroicity at the atomic scale 
will shed light on the effective magnetoelectric coupling mechanisms taking place at the nanometer scale.

The pantograph mechanism, which is the key ingredient in our proposal to generate the magnetoelectric coupling,  
encodes the relation between the dipolar moments and their lattice environment and 
is present as well in two or three dimensional systems. 
Appropriate extensions of the present model can be written taking into account detailed crystallographic data.
In these higher dimensional settings the isolated solitons could become extended walls; 
the predicted magnetically driven polarization switch  off  will probably survive to these generalizations.

\section*{Acknowledgements}

The authors are grateful to A.E. Trumper for early interest and comments on this work. D.C.C. acknowledges useful discussions with S.-W. Cheong, D.I. Khomskii and  C.D. Batista. 
This work was
partially supported by CONICET (Grants No. PIP 2015-813
and No. PIP 2015-364), Argentina.


\end{document}